# The Numerical Simulations and Resonant Scattering of Intensive Electromagnetic Fields of Waves by Dielectric Layer with Kerr-Like Nonlinearity

Vasyl V. Yatsyk*

Usikov Institute for Radiophysics & Electronics
National Academy of Sciences of Ukraine

## Abstract

On an example of the open nonlinear electrodynamic system - transverse non-homogeneous, isotropic, nonmagnetic, linearly polarized, nonlinear (a Kerr-like dielectric nonlinearity) dielectric layer, the algorithms of solution of the diffraction problem of a plane wave on the nonlinear object and the results of the numerical analysis of the nonlinear problem are shown. The nonlinear diffraction problem reduced the solutions of the systems of the non-homogeneous nonlinear equations of the second kind on the base of the iterative schemes. The results of the numerical analysis of the diffraction problem of a plane wave on the nonlinear object are shown. The effects: non-uniform shift of resonant frequency of the diffraction characteristics of a nonlinear dielectric layer; increase of the angle of the transparency of the nonlinear layer at growth of intensity of the field are found out.

# 1 The Diffraction Problem

Maxwell's equations:

$$\begin{aligned} &\nabla\times\vec{E}(\vec{r},t)=-\frac{1}{c}\frac{\partial\vec{B}(\vec{r},t)}{\partial t}, \quad \nabla\times\vec{H}(\vec{r},t)=\frac{1}{c}\frac{\partial\vec{D}(\vec{r},t)}{\partial t}, \\ &\nabla\cdot\vec{D}(\vec{r},t)=0, \qquad\qquad\quad \nabla\cdot\vec{B}(\vec{r},t)=0, \end{aligned} \tag{1}$$

and the material equations:

$$\begin{aligned} \vec{D}(\vec{r},t) &= \vec{E}(\vec{r},t)+4\pi\vec{P}(\vec{r},t), \\ \vec{B}(\vec{r},t) &= \vec{H}(\vec{r},t)+4\pi\vec{M}(\vec{r},t) \end{aligned}. \tag{2}$$

When $\vec{M}(\vec{r},t)=0$ the equations (1), (2) are reduced to (see [1]):

$$\nabla^2\vec{E}(\vec{r},t)-\nabla(\nabla\vec{E}(\vec{r},t))-\frac{1}{c^2}\frac{\partial^2}{\partial t^2}\vec{D}^{(L)}(\vec{r},t)-\frac{4\pi}{c^2}\frac{\partial^2}{\partial t^2}\vec{P}^{(NL)}(\vec{r},t)=0. \tag{3}$$

* E-mail address: yatsyk@vk.kharkov.ua, yatsyk@ire.kharkov.ua

Here: $\vec{D}^{(L)} = \vec{E} + 4\pi\vec{P}^{(L)} = \hat{\varepsilon}\vec{E}$ ; $\vec{P}^{(L)} = \hat{\chi}^{(1)}\vec{E}$ ; $\vec{D}_i^{(L)} = \varepsilon_{ij}^{(L)}\vec{E}_j$ ; $\varepsilon_{ij}^{(L)} = 1+4\pi\chi_{ij}^{(1)}$ ; $\vec{E}=(E_x,E_y,E_z)$; $\vec{P} = (P_x, P_y, P_z)$; $P_i \equiv P_i^{(L)} + P_i^{(NL)}$ ; $P_i^{(L)} \equiv \chi_{ij}^{(1)}E_j$ ; $P_i^{(NL)} \equiv \chi_{ijk}^{(2)}E_jE_k + \chi_{ijkl}^{(3)}E_jE_kE_l + \ldots$ ; $\varepsilon_{ij}^{(L)}$ are components of a tensor of a linear part of dielectric permittivity $\hat{\varepsilon}$ ; accordingly these parameters $\chi_{ij}^{(1)}$, $\chi_{ijk}^{(2)}$, $\chi_{ijkl}^{(3)}$, … are components of the appropriate tensors of susceptibilities $\hat{\chi}^{(1)}$, $\hat{\chi}^{(2)}$, $\hat{\chi}^{(3)}$, … .

Let $\vec{E}(\vec{r},t) = \exp(-i\omega t)\cdot\vec{E}(\vec{r})$. We consider a nonmagnetic $\vec{M} = 0$, isotropic, transverse non-homogeneous $\varepsilon^{(L)}(z) = \varepsilon_{xx}^{(L)}(z)$, linearly polarized $\vec{E} = (E_x, 0, 0)$, $\vec{H} = (0, H_y, H_z)$ (E-polarized) and Kerr-like nonlinearity $P_x^{(NL)} = (3/4)\chi_{xxxx}^{(3)}|E_x|^2 E_x$ (where $\vec{P}^{(NL)} = (P_x^{(NL)},\ 0, 0)$) dielectric layer (Fig. 1), [1-3].

In this case (see (1), (2) and [1]): $\nabla\cdot\vec{D}=0 \;\Rightarrow\; \nabla\cdot\vec{E} = -\left(\vec{E}\cdot(\nabla\cdot\hat{\varepsilon})/\hat{\varepsilon}\right) \;\Rightarrow\; \nabla\cdot(\nabla\cdot\vec{E}) = 0$.

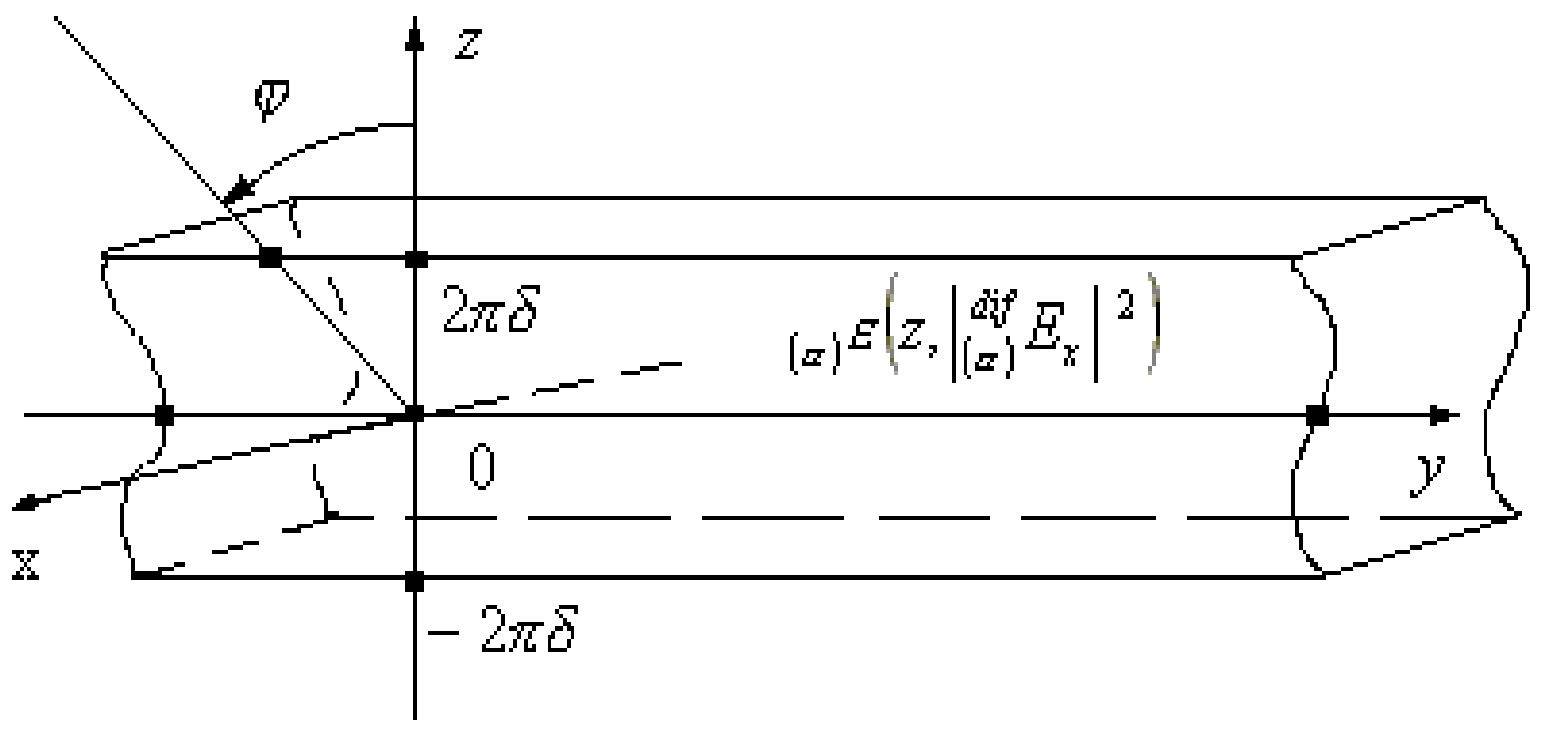

**Figure 1.** Nonlinear dielectric layer.

The complete diffraction field ${}^{dif}_{(\alpha)}E_x(y,z) = {}^{inc}E_x(y,z) + {}^{scat}_{(\alpha)}E_x(y,z)$ of a plane wave ${}^{inc}E_x(y,z) = {}^{inc}a\exp[i(\phi y - \Gamma\cdot(z-2\pi\delta))]$, $z > 2\pi\delta$ on the nonlinear dielectric layer (Fig. 1.) satisfies such conditions of the problem (see (3)):

$$\begin{aligned}&\Delta\cdot\vec{E} + \frac{\omega^2}{c^2}\cdot\varepsilon^{(L)}(z)\cdot\vec{E} + \frac{4\pi\omega^2}{c^2}\cdot\vec{P}^{(NL)} \equiv \\ &\equiv\left(\Delta + \kappa^2\cdot{}_{(\alpha)}\varepsilon\left(z,\left|{}^{dif}_{(\alpha)}E_x\right|^2\right)\right)\cdot{}^{dif}_{(\alpha)}E_x(y,z) = 0\end{aligned}, \tag{4}$$

the generalized boundary conditions:

$$\begin{aligned}&{}^{dif}_{(\alpha)}E_{tg} \text{ and } {}^{dif}_{(\alpha)}H_{tg} \text{ are continuous at discontinuities } {}_{(\alpha)}\varepsilon\left(z,\left|{}^{dif}_{(\alpha)}E_x\right|^2\right);\\ &{}^{dif}_{(\alpha)}E_x(y,z) = {}^{dif}_{(\alpha)}U(z)\cdot\exp(i\phi y)\text{, the condition of spatial quasihomogeneity on } y;\end{aligned} \tag{5}$$

the condition of the radiation for scattered field:

$$ {}^{scat}_{(\alpha)}E_x(y,z) = \left\{ \begin{matrix} {}^{scat}_{(\alpha)}a \\ {}^{scat}_{(\alpha)}b \end{matrix} \right\} \cdot \exp\left(i \cdot \left(\phi y \pm \Gamma \cdot \left(z \mp 2\pi\delta\right)\right)\right), \quad z \begin{matrix} > \\ < \end{matrix} \pm 2\pi\delta \qquad (6) $$

Here: ${}_{(\alpha)}\varepsilon\left(z, \left|{}^{dif}_{(\alpha)}E_x\right|^2\right) = \begin{cases} 1, & |z| > 2\pi\delta \\ \varepsilon^{(L)}(z) + \alpha \cdot \left|{}^{dif}_{(\alpha)}E_x\right|^2, & |z| \le 2\pi\delta \end{cases}$; $\Delta = \partial^2/\partial y^2 + \partial^2/\partial z^2$; $\alpha = 3\pi\chi^{(3)}_{xxxx}$; $\Gamma = (\kappa^2 - \phi^2)^{1/2}$; $\phi \equiv \kappa \cdot \sin(\varphi)$; $|\varphi| < \pi/2$ (see Fig. 1); $\kappa = \omega/c \equiv 2\pi/\lambda$; $c = (\varepsilon_0 \mu_0)^{-1/2}$, $\varepsilon_0$, $\mu_0$ and $\lambda$ length of the wave is the parameters of environment.

The required of the solution of problem (4)-(6) are of kind:

$$ {}^{dif}_{(\alpha)}E_x(y,z) = {}^{dif}_{(\alpha)}U(z) \cdot \exp(i\,\phi\, y) = $$
$$ = \begin{cases} {}^{inc}a \cdot \exp\left(i \cdot \left(\phi\, y - \Gamma \cdot (z - 2\pi\delta)\right)\right) + {}^{scat}_{(\alpha)}a \cdot \exp\left(i \cdot \left(\phi\, y + \Gamma \cdot (z - 2\pi\delta)\right)\right), & z > 2\pi\delta, \\ {}^{scat}_{(\alpha)}U(z) \cdot \exp(i \cdot \phi\, y), & |z| \le 2\pi\delta, \\ {}^{scat}_{(\alpha)}b \cdot \exp\left(i \cdot \left(\phi\, y - \Gamma \cdot (z + 2\pi\delta)\right)\right), & z < -2\pi\delta. \end{cases} \qquad (7) $$

Here ${}^{dif}_{(\alpha)}U(-2\pi\delta) = {}^{scat}_{(\alpha)}b$, ${}^{dif}_{(\alpha)}U(2\pi\delta) = {}^{inc}a + {}^{scat}_{(\alpha)}a$.

The nonlinear problem (4)-(6) is reduced to finding the solutions ${}^{dif}_{(\alpha)}U(z) \in L_2([-2\pi\delta, 2\pi\delta])$ (see (7)) of the non-homogeneous nonlinear integrated equation of the second kind [4, 5]:

$$ {}^{dif}_{(\alpha)}U(z) + \frac{i\kappa^2}{2\Gamma} \int_{-2\pi\delta}^{2\pi\delta} \exp\left(i\Gamma \cdot |z - z_0|\right) \cdot \left[1 - \left(\varepsilon^{(L)}(z_0) + \alpha \left|{}^{dif}_{(\alpha)}U(z_0)\right|^2\right)\right] \cdot {}^{dif}_{(\alpha)}U(z_0)\, dz_0 = {}^{inc}U(z), \quad |z| \le 2\pi\delta, \qquad (8) $$

where ${}^{inc}U(z) = {}^{inc}a \exp\left[-i\Gamma \cdot (z - 2\pi\delta)\right]$.

# 2 Numerical Simulations

## 2.1 Iteration Scheme on the Basis of Simpson's Quadrature

The integrated equation (8) with application of the quadrature method is reduced to the system of the nonlinear equations of the second kind [6-11]:

$$ (E - {}_{(\alpha)}B(\left|{}^{dif}_{(\alpha)}U\right|^2)) \cdot {}^{dif}_{(\alpha)}U = {}^{inc}U \qquad (9) $$

Here: $z_1 = -2\pi\delta < z_2 < \ldots < z_n < \ldots < z_N = 2\pi\delta$; ${}_{(\alpha)}B\left(\left|{}^{dif}_{(\alpha)}U\right|^2\right) = \left\{A_m \cdot {}_{(\alpha)}K_{nm}\left(\left|{}^{dif}_{(\alpha)}U\right|^2\right)\right\}_{n,m=1}^{N}$; ${}_{(\alpha)}K_{nm}\left(\left|{}^{dif}_{(\alpha)}U\right|^2\right) = \left(i\kappa^2/(2\Gamma)\right) \cdot \exp\left(i \cdot \Gamma \cdot |z_n - z_m|\right) \cdot \left[1 - \left(\varepsilon^{(L)}(z_m) + \alpha \cdot \left|{}^{dif}_{(\alpha)}U(z_m)\right|^2\right)\right]$; ${}^{dif}_{(\alpha)}U = \left\{{}^{dif}_{(\alpha)}U(z_n)\right\}_{n=1}^{N}$; $E = \left\{\delta_n^m\right\}_{n,m=1}^{N}$; ${}^{inc}U = \left\{{}^{inc}a \cdot \exp\left[-i \cdot \Gamma \cdot (z_n - 2\pi\delta)\right]\right\}_{n=1}^{N}$; $\delta_n^m$ is the Kronecker delta; $A_m$ are the numerical coefficients dictated by chosen quadrature form.

Solutions of the system (9) are carried out by the method of iterations:

$$\left\{\left(E - {}_{(\alpha)}B\left(\left|{}^{dif}_{(\alpha)}U^{(s-1)}\right|^2\right)\right) \cdot {}^{dif}_{(\alpha)}U^{(s)} = {}^{inc}U\right\}_{s=1}^{S:\left\|{}^{dif}_{(\alpha)}U^{(s)} - {}^{dif}_{(\alpha)}U^{(s-1)}\right\| / \left\|{}^{dif}_{(\alpha)}U^{(s)}\right\| < \xi} . \quad (10)$$

Here index $\xi$ is the given meaning of a relative error.

## 2.2 Newtonian Algorithm and Simpson's Quadrature

Integrand function in (8) is not holomorphic on ${}^{dif}_{(\alpha)}U(z)$, but it is holomorphic in space of arguments $\left\{\mathrm{Re}\left({}^{dif}_{(\alpha)}U(z)\right), \mathrm{Im}\left({}^{dif}_{(\alpha)}U(z)\right)\right\}$. Applying a method of quadratures in space of real values of arguments $\left\{{}^{dif}_{(\alpha)}U' \equiv \mathrm{Re}\left({}^{dif}_{(\alpha)}U\right), {}^{dif}_{(\alpha)}U'' \equiv \mathrm{Im}\left({}^{dif}_{(\alpha)}U\right)\right\}$ we have [9, 11], see (8) and (9):

$$\left(E - {}_{(\alpha)}B\left(\left\{{}^{dif}_{(\alpha)}U'\right\}^2 + \left\{{}^{dif}_{(\alpha)}U''\right\}^2\right)\right)\left({}^{dif}_{(\alpha)}U' + i\,{}^{dif}_{(\alpha)}U''\right) - {}^{inc}U = 0 \quad (11)$$

Here: $z_1 = -2\pi\delta < z_2 < \ldots < z_n < \ldots < z_N = 2\pi\delta$; ${}^{inc}U = \left\{{}^{inc}a \cdot \exp\left[-i\Gamma\cdot\left(z_n - 2\pi\delta\right)\right]\right\}_{n=1}^{N}$; $E = \left\{\delta_n^m\right\}_{n,m=1}^{N}$; $\delta_n^m$ is the Kronecker delta; ${}^{dif}_{(\alpha)}U' = \left\{{}^{dif}_{(\alpha)}U'(z_n) \equiv {}^{dif}_{(\alpha)}U'_n\right\}_{n=1}^{N}$; ${}^{dif}_{(\alpha)}U'' = \left\{{}^{dif}_{(\alpha)}U''(z_n) \equiv {}^{dif}_{(\alpha)}U''_n\right\}_{n=1}^{N}$; ${}_{(\alpha)}B\left(\left\{{}^{dif}_{(\alpha)}U'\right\}^2 + \left\{{}^{dif}_{(\alpha)}U''\right\}^2\right) = \left\{A_m \cdot {}_{(\alpha)}K_{nm}\left(\left\{{}^{dif}_{(\alpha)}U'\right\}^2 + \left\{{}^{dif}_{(\alpha)}U''\right\}^2\right)\right\}_{n,m=1}^{N}$; ${}_{(\alpha)}K_{nm}\left(\left\{{}^{dif}_{(\alpha)}U'\right\}^2 + \left\{{}^{dif}_{(\alpha)}U''\right\}^2\right) = \frac{i\kappa^2}{2\Gamma}\exp\left(i\Gamma\cdot|z_n - z_m|\right)\cdot\left[1 - \left(\varepsilon^{(L)}(z_m) + \alpha\left(\left\{{}^{dif}_{(\alpha)}U'_m\right\}^2 + \left\{{}^{dif}_{(\alpha)}U''_m\right\}^2\right)\right)\right]$;
$A_m$ are the numerical coefficients dictated by chosen quadrature form.

System of the nonlinear equations (11) we shall write down as:

$$\left\{\begin{aligned} & f_n\left({}^{dif}_{(\alpha)}U'_1, {}^{dif}_{(\alpha)}U''_1, \ldots, {}^{dif}_{(\alpha)}U'_N, {}^{dif}_{(\alpha)}U''_N\right) \equiv {}^{dif}_{(\alpha)}U'_n + i\cdot{}^{dif}_{(\alpha)}U''_n - {}^{inc}a\cdot\exp\left[-i\Gamma\cdot\left(z_n - 2\pi\delta\right)\right] - \frac{i\kappa^2}{2\Gamma}\times \\ & \times\sum_{m=1}^{N} A_m\cdot\exp\left(i\Gamma\cdot|z_n - z_m|\right)\cdot\left[1 - \left(\varepsilon^{(L)}(z_m) + \alpha\left(\left\{{}^{dif}_{(\alpha)}U'_m\right\}^2 + \left\{{}^{dif}_{(\alpha)}U''_m\right\}^2\right)\right)\right]\left({}^{dif}_{(\alpha)}U'_m + i\,{}^{dif}_{(\alpha)}U''_m\right) = 0 \end{aligned}\right\}_{n=1}^{N} . \quad (12)$$

Let's write down (12) in a vector kind:

$$\left\{\begin{aligned} \mathrm{Re} f_n\left({}^{dif}_{(\alpha)}U'_1, {}^{dif}_{(\alpha)}U''_1, \ldots, {}^{dif}_{(\alpha)}U'_N, {}^{dif}_{(\alpha)}U''_N\right) = 0 \\ \mathrm{Im} f_n\left({}^{dif}_{(\alpha)}U'_1, {}^{dif}_{(\alpha)}U''_1, \ldots, {}^{dif}_{(\alpha)}U'_N, {}^{dif}_{(\alpha)}U''_N\right) = 0 \end{aligned}\right\}_{n=1}^{N} \Leftrightarrow \vec{f}\left({}^{dif}_{(\alpha)}\vec{U}\right) = \vec{0}, \quad (13)$$

here ${}^{dif}_{(\alpha)}\vec{U} \equiv \left({}^{dif}_{(\alpha)}U'_1, {}^{dif}_{(\alpha)}U''_1, \ldots, {}^{dif}_{(\alpha)}U'_p, {}^{dif}_{(\alpha)}U''_p, \ldots, {}^{dif}_{(\alpha)}U'_N, {}^{dif}_{(\alpha)}U''_N\right)^T$, $T$ - transposition. The solution (13) could be found by a Newtonian method [12]:

$$J^k \cdot \left({}^{dif}_{(\alpha)}\vec{U}^{k+1} - {}^{dif}_{(\alpha)}\vec{U}^{k}\right) = -\vec{f}\left({}^{dif}_{(\alpha)}\vec{U}^{k}\right), \quad \text{when} \quad k \to \infty, \quad {}^{dif}_{(\alpha)}\vec{U}^{k} \to {}^{dif}_{(\alpha)}\vec{U}, \quad (14)$$

here $J^k \equiv J\left(\overset{dif}{(\alpha)}\vec{U}^k\right)$ is the $k$-th iteration of the Jacobi matrixes $J\left(\overset{dif}{(\alpha)}\vec{U}\right)$ for vector function $\vec{f}\left(\overset{dif}{(\alpha)}\vec{U}\right)$:

$$\vec{f}\left(\overset{dif}{(\alpha)}\vec{U}\right)\equiv\begin{pmatrix} \mathrm{Re} f_1\left(\overset{dif}{(\alpha)}\vec{U}\right) \\ \mathrm{Im} f_1\left(\overset{dif}{(\alpha)}\vec{U}\right) \\ \ldots\ldots\ldots \\ \ldots\ldots\ldots \\ \ldots\ldots\ldots \\ \ldots\ldots\ldots \\ \mathrm{Re} f_N\left(\overset{dif}{(\alpha)}\vec{U}\right) \\ \mathrm{Im} f_N\left(\overset{dif}{(\alpha)}\vec{U}\right) \end{pmatrix},\quad J\left(\overset{dif}{(\alpha)}\vec{U}\right)\equiv\begin{pmatrix} \mathrm{Re}\frac{\partial f_1\left(\overset{dif}{(\alpha)}\vec{U}\right)}{\partial\overset{dif}{(\alpha)}U'_1} & \mathrm{Re}\frac{\partial f_1\left(\overset{dif}{(\alpha)}\vec{U}\right)}{\partial\overset{dif}{(\alpha)}U''_1} & \ldots & \mathrm{Re}\frac{\partial f_1\left(\overset{dif}{(\alpha)}\vec{U}\right)}{\partial\overset{dif}{(\alpha)}U'_N} & \mathrm{Re}\frac{\partial f_1\left(\overset{dif}{(\alpha)}\vec{U}\right)}{\partial\overset{dif}{(\alpha)}U''_N} \\ \mathrm{Im}\frac{\partial f_1\left(\overset{dif}{(\alpha)}\vec{U}\right)}{\partial\overset{dif}{(\alpha)}U'_1} & \mathrm{Im}\frac{\partial f_1\left(\overset{dif}{(\alpha)}\vec{U}\right)}{\partial\overset{dif}{(\alpha)}U''_1} & \ldots & \mathrm{Im}\frac{\partial f_1\left(\overset{dif}{(\alpha)}\vec{U}\right)}{\partial\overset{dif}{(\alpha)}U'_N} & \mathrm{Im}\frac{\partial f_1\left(\overset{dif}{(\alpha)}\vec{U}\right)}{\partial\overset{dif}{(\alpha)}U''_N} \\ \ldots & \ldots & \ldots & \ldots & \ldots \\ \mathrm{Re}\frac{\partial f_N\left(\overset{dif}{(\alpha)}\vec{U}\right)}{\partial\overset{dif}{(\alpha)}U'_1} & \mathrm{Re}\frac{\partial f_N\left(\overset{dif}{(\alpha)}\vec{U}\right)}{\partial\overset{dif}{(\alpha)}U''_1} & \ldots & \mathrm{Re}\frac{\partial f_N\left(\overset{dif}{(\alpha)}\vec{U}\right)}{\partial\overset{dif}{(\alpha)}U'_N} & \mathrm{Re}\frac{\partial f_N\left(\overset{dif}{(\alpha)}\vec{U}\right)}{\partial\overset{dif}{(\alpha)}U''_N} \\ \mathrm{Im}\frac{\partial f_N\left(\overset{dif}{(\alpha)}\vec{U}\right)}{\partial\overset{dif}{(\alpha)}U'_1} & \mathrm{Im}\frac{\partial f_N\left(\overset{dif}{(\alpha)}\vec{U}\right)}{\partial\overset{dif}{(\alpha)}U''_1} & \ldots & \mathrm{Im}\frac{\partial f_N\left(\overset{dif}{(\alpha)}\vec{U}\right)}{\partial\overset{dif}{(\alpha)}U'_N} & \mathrm{Im}\frac{\partial f_N\left(\overset{dif}{(\alpha)}\vec{U}\right)}{\partial\overset{dif}{(\alpha)}U''_N} \end{pmatrix} \quad (15)$$

at the following functions (for $n, p = 1, 2, \ldots, N$), see (12):

$$\begin{aligned} & f_n\left(\overset{dif}{(\alpha)}\vec{U}\right)=\overset{dif}{(\alpha)}U'_n + i\cdot\overset{dif}{(\alpha)}U''_n - {}^{inc}a\cdot\exp[-i\cdot\Gamma\cdot(z_n-2\pi\delta)]-\frac{i\kappa^2}{2\Gamma}\times \\ & \times\sum_{m=1}^{N}A_m\cdot\exp(i\Gamma\cdot|z_n-z_m|)\cdot\left[1-\left(\varepsilon^{(L)}(z_m)+\alpha\cdot\left(\left\{\overset{dif}{(\alpha)}U'_m\right\}^2+\left\{\overset{dif}{(\alpha)}U''_m\right\}^2\right)\right)\right]\cdot\left(\overset{dif}{(\alpha)}U'_m + i\overset{dif}{(\alpha)}U''_m\right); \\ & \frac{\partial f_n\left(\overset{dif}{(\alpha)}\vec{U}\right)}{\partial\overset{dif}{(\alpha)}U'_p}=\delta_p^n-\frac{i\kappa^2}{2\Gamma}\cdot A_p\cdot\exp(i\cdot\Gamma\cdot|z_n-z_p|)\times \\ & \times\left[1-\varepsilon^{(L)}(z_p)-\alpha\cdot\left(3\left\{\overset{dif}{(\alpha)}U'_p\right\}^2+2i\cdot\overset{dif}{(\alpha)}U'_p\cdot\overset{dif}{(\alpha)}U''_p+\left\{\overset{dif}{(\alpha)}U''_p\right\}^2\right)\right]; \\ & \frac{\partial f_n\left(\overset{dif}{(\alpha)}\vec{U}\right)}{\partial\overset{dif}{(\alpha)}U''_p}=i\cdot\delta_p^n+\frac{\kappa^2}{2\Gamma}\cdot A_p\cdot\exp(i\cdot\Gamma\cdot|z_n-z_p|)\times \\ & \times\left[1-\varepsilon^{(L)}(z_p)-\alpha\cdot\left(3\left\{\overset{dif}{(\alpha)}U''_p\right\}^2-2i\cdot\overset{dif}{(\alpha)}U'_p\cdot\overset{dif}{(\alpha)}U''_p+\left\{\overset{dif}{(\alpha)}U'_p\right\}^2\right)\right]. \end{aligned} \quad (16)$$

## 2.3 Newtonian Algorithm and Teilor-Series of the Diffraction Field

Let's find the solution of the equation (8) as:

$$\overset{dif}{(\alpha)}U(z)=\sum_{n=0}^{L}\overset{dif}{(\alpha)}c_n\cdot z^n=\sum_{n=0}^{L}\left(\overset{dif}{(\alpha)}c'_n+i\overset{dif}{(\alpha)}c''_n\right)\cdot z^n,\quad \overset{dif}{(\alpha)}c'_n\equiv\mathrm{Re}\overset{dif}{(\alpha)}c_n,\ \overset{dif}{(\alpha)}c''_n\equiv\mathrm{Im}\overset{dif}{(\alpha)}c_n\in R;\ |z|\le 2\pi\delta. \quad (17)$$

After substitution (17) in (8), calculations of value of the equation (8) and its derivatives $\left\{\partial^l/\partial z^l\right\}_{l=1}^{L}$ in the point $z=0$, we obtain the system of differentiable functions in space of arguments $\left\{\overset{dif}{(\alpha)}c'_n,\ \overset{dif}{(\alpha)}c''_n\right\}_{n=0}^{L}$, see [6, 7]:

$$\left\{\begin{array}{l} f_l\left({}^{dif}_{(\alpha)}c'_0,\ {}^{dif}_{(\alpha)}c''_0,\dots,\ {}^{dif}_{(\alpha)}c'_L,\ {}^{dif}_{(\alpha)}c''_L\right) = {}^{dif}_{(\alpha)}c'_l + i\cdot{}^{dif}_{(\alpha)}c''_l - {}^{inc}a\dfrac{(-i\Gamma)^l}{\delta_l^0+l}\exp(i\Gamma 2\pi\delta) + \kappa^2\dfrac{(-i\Gamma)^{l-1}}{\delta_l^0+l}\times \\ \times\displaystyle\int_{-2\pi\delta}^{2\pi\delta} sign\left(z_0{}^l\right)\cdot\exp\left(i\cdot\Gamma\cdot|z_0|\right)\cdot\left[1-\varepsilon^{(L)}(z_0)-\alpha\left(\left\{\sum_{n=0}^L {}^{dif}_{(\alpha)}c'_n\cdot z_0{}^n\right\}^2+\left\{\sum_{n=0}^L {}^{dif}_{(\alpha)}c''_n\cdot z_0{}^n\right\}^2\right)\right]\times \\ \times\left(\displaystyle\sum_{n=0}^L\left({}^{dif}_{(\alpha)}c'_n+i\cdot{}^{dif}_{(\alpha)}c''_n\right)\cdot z_0{}^n\right)dz_0 = 0 \end{array}\right\}_{l=0}^{L} \tag{18}$$

Let's write down system (18) in a vector kind:

$$\left\{\begin{array}{l} \mathrm{Re}\, f_l\left({}^{dif}_{(\alpha)}c'_0,\ {}^{dif}_{(\alpha)}c''_0,\dots,\ {}^{dif}_{(\alpha)}c'_L,\ {}^{dif}_{(\alpha)}c''_L\right)=0 \\ \mathrm{Im}\, f_l\left({}^{dif}_{(\alpha)}c'_0,\ {}^{dif}_{(\alpha)}c''_0,\dots,\ {}^{dif}_{(\alpha)}c'_L,\ {}^{dif}_{(\alpha)}c''_L\right)=0 \end{array}\right\}_{l=0}^{L} \Leftrightarrow \vec{f}\left({}^{dif}_{(\alpha)}\vec{c}\right)=\vec{0}, \tag{19}$$

here: ${}^{dif}_{(\alpha)}\vec{c} \equiv \left({}^{dif}_{(\alpha)}c'_0,\ {}^{dif}_{(\alpha)}c''_0,\dots,\ {}^{dif}_{(\alpha)}c'_L,\ {}^{dif}_{(\alpha)}c''_L\right)^T$, $\vec{f}\left({}^{dif}_{(\alpha)}\vec{c}\right) \equiv \left(\mathrm{Re} f_0\left({}^{dif}_{(\alpha)}\vec{c}\right), \mathrm{Im} f_0\left({}^{dif}_{(\alpha)}\vec{c}\right),\dots,\mathrm{Re} f_L\left({}^{dif}_{(\alpha)}\vec{c}\right), \mathrm{Im} f_L\left({}^{dif}_{(\alpha)}\vec{c}\right)\right)^T$.

The solution (19) could be found by a method of Newton (see [12]):

$$J^k\cdot\left({}^{dif}_{(\alpha)}\vec{c}^{\,k+1} - {}^{dif}_{(\alpha)}\vec{c}^{\,k}\right) = -\vec{f}\left({}^{dif}_{(\alpha)}\vec{c}^{\,k}\right), \quad \text{when } k\to\infty,\ {}^{dif}_{(\alpha)}\vec{c}^{\,k}\to{}^{dif}_{(\alpha)}\vec{c}, \tag{20}$$

here $J^k \equiv J\left({}^{dif}_{(\alpha)}\vec{c}^{\,k}\right)$ is the $k$-th iteration of the Jacobi matrixes $J\left({}^{dif}_{(\alpha)}\vec{c}\right)$ for vector function $\vec{f}\left({}^{dif}_{(\alpha)}\vec{c}\right)$.

Let's note, that having presented a linear part of dielectric permeability as

$$\varepsilon^{(L)}(z) = \sum_{g=0}^{G}\varepsilon_g\cdot z^g\ ,\ |z| \le 2\pi\delta\ , \tag{21}$$

here $\{\varepsilon_g\}_{g=0}^{G} \in C$, $C$ is the complex set. We easily receive obvious, analytical expressions for the function $\vec{f}\left({}^{dif}_{(\alpha)}\vec{c}\right)$ and Jacobi matrix $J\left({}^{dif}_{(\alpha)}\vec{c}\right)$:

$$\vec{f}\left({}^{dif}_{(\alpha)}\vec{c}\right) \equiv \begin{pmatrix} \mathrm{Re} f_0\left({}^{dif}_{(\alpha)}\vec{c}\right) \\ \mathrm{Im} f_0\left({}^{dif}_{(\alpha)}\vec{c}\right) \\ \dots\dots \\ \dots\dots \\ \dots\dots \\ \dots\dots \\ \mathrm{Re} f_L\left({}^{dif}_{(\alpha)}\vec{c}\right) \\ \mathrm{Im} f_L\left({}^{dif}_{(\alpha)}\vec{c}\right) \end{pmatrix},\ J\left({}^{dif}_{(\alpha)}\vec{c}\right) \equiv \begin{pmatrix} \mathrm{Re}\dfrac{\partial f_0\left({}^{dif}_{(\alpha)}\vec{c}\right)}{\partial\,{}^{dif}_{(\alpha)}c'_0} & \mathrm{Re}\dfrac{\partial f_0\left({}^{dif}_{(\alpha)}\vec{c}\right)}{\partial\,{}^{dif}_{(\alpha)}c''_0} & \dots & \mathrm{Re}\dfrac{\partial f_0\left({}^{dif}_{(\alpha)}\vec{c}\right)}{\partial\,{}^{dif}_{(\alpha)}c'_L} & \mathrm{Re}\dfrac{\partial f_0\left({}^{dif}_{(\alpha)}\vec{c}\right)}{\partial\,{}^{dif}_{(\alpha)}c''_L} \\ \mathrm{Im}\dfrac{\partial f_0\left({}^{dif}_{(\alpha)}\vec{c}\right)}{\partial\,{}^{dif}_{(\alpha)}c'_0} & \mathrm{Im}\dfrac{\partial f_0\left({}^{dif}_{(\alpha)}\vec{c}\right)}{\partial\,{}^{dif}_{(\alpha)}c''_0} & \dots & \mathrm{Im}\dfrac{\partial f_0\left({}^{dif}_{(\alpha)}\vec{c}\right)}{\partial\,{}^{dif}_{(\alpha)}c'_L} & \mathrm{Im}\dfrac{\partial f_0\left({}^{dif}_{(\alpha)}\vec{c}\right)}{\partial\,{}^{dif}_{(\alpha)}c''_L} \\ \dots & \dots & \dots & \dots & \dots \\ \mathrm{Re}\dfrac{\partial f_L\left({}^{dif}_{(\alpha)}\vec{c}\right)}{\partial\,{}^{dif}_{(\alpha)}c'_0} & \mathrm{Re}\dfrac{\partial f_L\left({}^{dif}_{(\alpha)}\vec{c}\right)}{\partial\,{}^{dif}_{(\alpha)}c''_0} & \dots & \mathrm{Re}\dfrac{\partial f_L\left({}^{dif}_{(\alpha)}\vec{c}\right)}{\partial\,{}^{dif}_{(\alpha)}c'_L} & \mathrm{Re}\dfrac{\partial f_L\left({}^{dif}_{(\alpha)}\vec{c}\right)}{\partial\,{}^{dif}_{(\alpha)}c''_L} \\ \mathrm{Im}\dfrac{\partial f_L\left({}^{dif}_{(\alpha)}\vec{c}\right)}{\partial\,{}^{dif}_{(\alpha)}c'_0} & \mathrm{Im}\dfrac{\partial f_L\left({}^{dif}_{(\alpha)}\vec{c}\right)}{\partial\,{}^{dif}_{(\alpha)}c''_0} & \dots & \mathrm{Im}\dfrac{\partial f_L\left({}^{dif}_{(\alpha)}\vec{c}\right)}{\partial\,{}^{dif}_{(\alpha)}c'_L} & \mathrm{Im}\dfrac{\partial f_L\left({}^{dif}_{(\alpha)}\vec{c}\right)}{\partial\,{}^{dif}_{(\alpha)}c''_L} \end{pmatrix} \tag{22}$$

at the following functions (for $l, p = 0, 1, 2, \dots, L$), see (18):

$$
\begin{aligned}
& f_l\left({}^{dif}_{(\alpha)}\vec{c}\right)= {}^{dif}_{(\alpha)}c'_l + i\,{}^{dif}_{(\alpha)}c''_l - {}^{inc}a\,\frac{(-i\Gamma)^l}{\delta_l^0+l}\exp\left(i\Gamma 2\pi\delta\right)+\kappa^2\frac{(-i\Gamma)^{l-1}}{\delta_l^0+l}\sum_{s=0}^{L}\left({}^{dif}_{(\alpha)}c'_s + i\cdot{}^{dif}_{(\alpha)}c''_s\right)\times \\
& \times\left\{V(l,s)-\sum_{g=0}^{G}\varepsilon_g V(l,s+g)-\alpha\sum_{n=0}^{L}\sum_{m=0}^{L}\left({}^{dif}_{(\alpha)}c'_n\cdot{}^{dif}_{(\alpha)}c'_m+{}^{dif}_{(\alpha)}c''_n\cdot{}^{dif}_{(\alpha)}c''_m\right)\cdot V(l,n+m+s)\right\}; \\
& \frac{\partial f_l\left({}^{dif}_{(\alpha)}\vec{c}\right)}{\partial\,{}^{dif}_{(\alpha)}c'_p}=\delta_p^l+\kappa^2\frac{(-i\cdot\Gamma)^{l-1}}{\delta_l^0+l}\cdot\left[V(l,p)-\sum_{g=0}^{G}\varepsilon_g\cdot V(l,p+g)-\right. \\
& \left.-\alpha\sum_{n=0}^{L}\sum_{m=0}^{L}\left(3\,{}^{dif}_{(\alpha)}c'_n\cdot{}^{dif}_{(\alpha)}c'_m+{}^{dif}_{(\alpha)}c''_n\cdot{}^{dif}_{(\alpha)}c''_m+2i\,{}^{dif}_{(\alpha)}c'_n\cdot{}^{dif}_{(\alpha)}c''_m\right)\cdot V(l,n+m+p)\right]; \\
& \frac{\partial f_l\left({}^{dif}_{(\alpha)}\vec{c}\right)}{\partial\,{}^{dif}_{(\alpha)}c''_p}=i\cdot\delta_p^l+i\cdot\kappa^2\frac{(-i\cdot\Gamma)^{l-1}}{\delta_l^0+l}\cdot\left[V(l,p)-\sum_{g=0}^{G}\varepsilon_g\cdot V(l,p+g)-\right. \\
& \left.-\alpha\sum_{n=0}^{L}\sum_{m=0}^{L}\left({}^{dif}_{(\alpha)}c'_n\cdot{}^{dif}_{(\alpha)}c'_m+3\,{}^{dif}_{(\alpha)}c''_n\cdot{}^{dif}_{(\alpha)}c''_m-2i\,{}^{dif}_{(\alpha)}c'_n\cdot{}^{dif}_{(\alpha)}c''_m\right)\cdot V(l,n+m+p)\right].
\end{aligned}
\tag{23}
$$

Here $V(l,p)\equiv\int_{-2\pi\delta}^{2\pi\delta}sign\left(z_0{}^l\right)\cdot\exp\left(i\cdot\Gamma\cdot|z_0|\right)\cdot z_0{}^p\,dz_0=\left(1+(-1)^{l+p}\right)\cdot W(p)$ and $W(p)=\exp(i\cdot\Gamma\cdot 2\pi\delta)\times$
$\times\left[\frac{(2\pi\delta)^p}{i\cdot\Gamma}+\left(1-\delta_p^0\right)\sum_{k=1}^{p}(-1)^k\frac{p(p-1)\cdot\ldots\cdot(p-k+1)}{(i\cdot\Gamma)^{k+1}}(2\pi\delta)^{p-k}\right]-\frac{\delta_p^0}{i\cdot\Gamma}-\left(1-\delta_p^0\right)(-1)^p\frac{p(p-1)\cdot\ldots\cdot 1}{(i\cdot\Gamma)^{p+1}}$.

# 3 The numerical analysis

## 3.1 Intensity And Resonant Frequency

The effect of non-uniform shift of resonant frequency of the diffraction characteristics of nonlinear dielectric layer is found out at increase of intensity of inciting field [8-11] (see Fig. 2, and also Fig. 3).

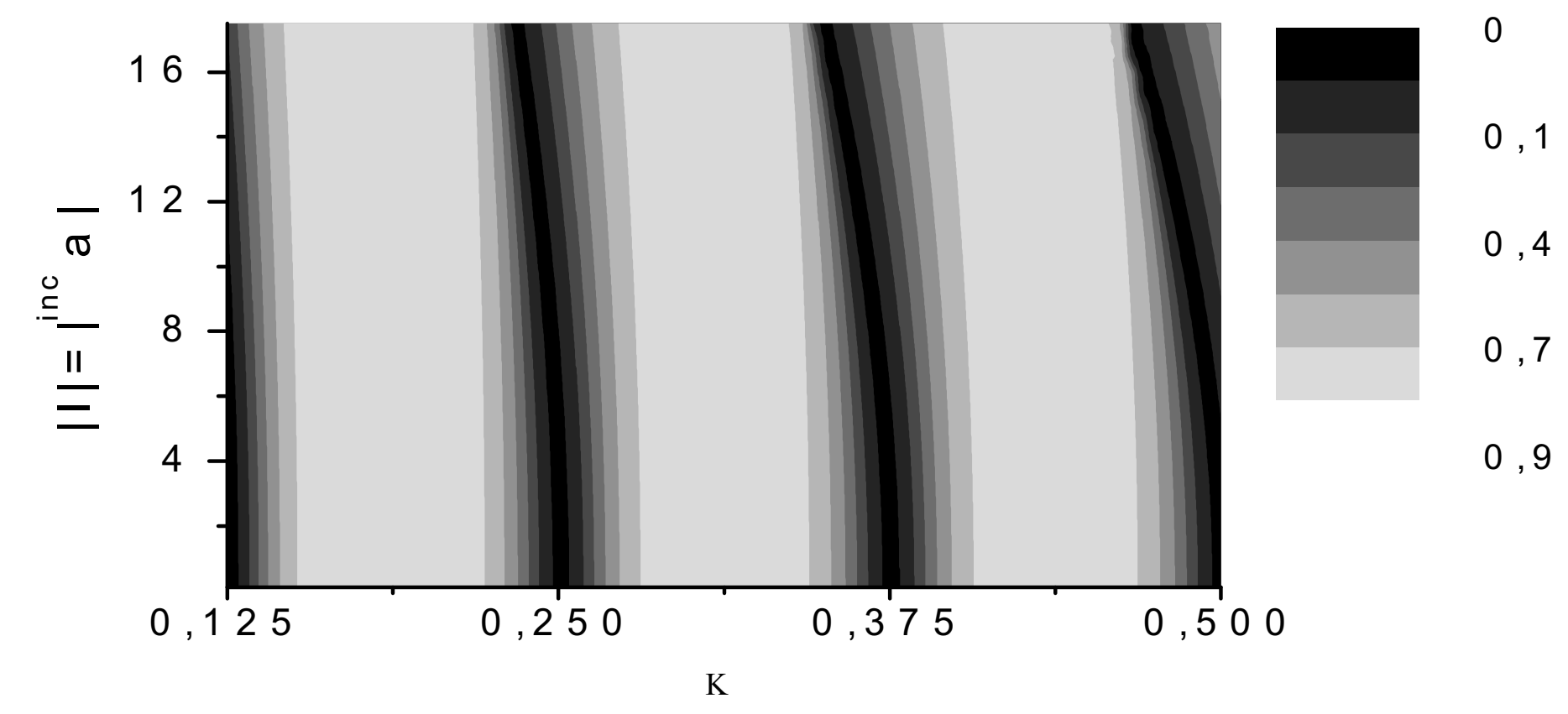

Share of the reflected wave: $\eta\left(|R(\alpha)|^2\right)$.

**Figure 2.** Parameters of the nonlinear problem: $\alpha=0{,}01$; $\varepsilon^{(L)}=16$; $\delta=0{,}5$; $\varphi=45^0$.

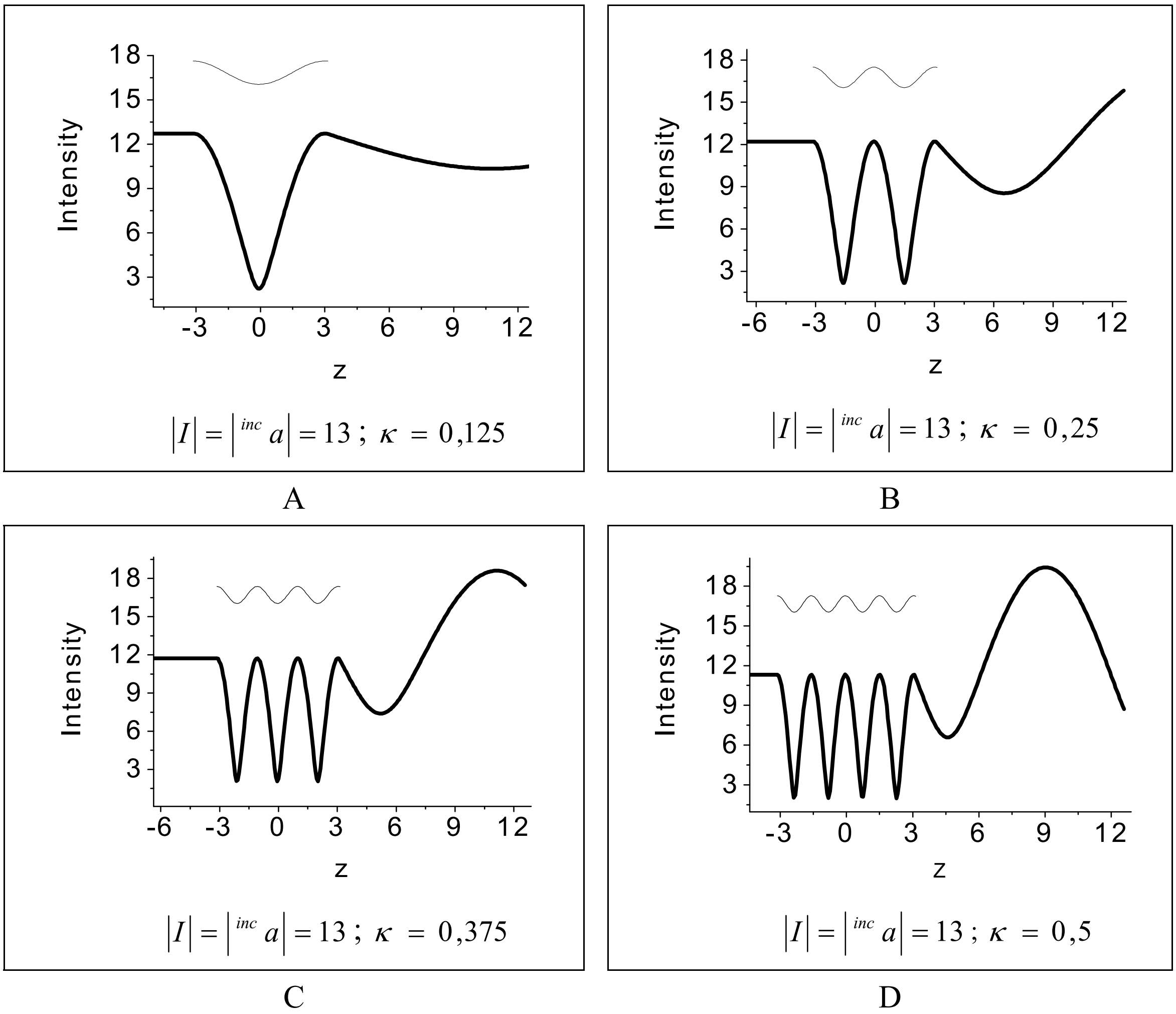

**Figure 3.** Parameters of structure and designations:

$\alpha = 0{,}01$; $\varepsilon^{(L)} = 16$; $\delta = 0{,}5$; $\varphi = 45^0$; ━━ $\left|{}^{dif}_{(\alpha)}E_x\right|$; —— ${}_{(\alpha)}\varepsilon\left(z, \left|{}^{dif}_{(\alpha)}E_x\right|^2\right)$.

Growth of intensity of the inciting field $|I| = \left|{}^{inc}a\right|$ results in change $\eta\left(|R(\alpha)|^2\right) = |R(\alpha)|^2 / |I|^2$: reduction of value of resonant frequency with increase and reduction of a steepness of the diffraction characteristics before and after resonant frequency (Fig. 2).

Here: $|R(\alpha)| \equiv \left|{}^{scat}_{(\alpha)}a\right|$, $|T(\alpha)| \equiv \left|{}^{scat}_{(\alpha)}b\right|$ and $|I|^2 = |T(\alpha)|^2 + |R(\alpha)|^2$.

## 3.2 Intensity and Transparency

The effect of increase of the angle of the transparency of the nonlinear layer ($\alpha \neq 0$) at growth of intensity of the inciting field is found out, [8, 10, 11]. See contour of coefficient of reflection on Fig. 4, A and graphics on Fig. 4, B for $\left|{}^{inc}a\right| = 4$ at $\varphi \approx 23^0$, Fig. 4, C for $\left|{}^{inc}a\right| = 8$ at $\varphi \approx 46^0$ and Fig. 4, D for $\left|{}^{inc}a\right| = 11$ at $\varphi \approx 76^0$.

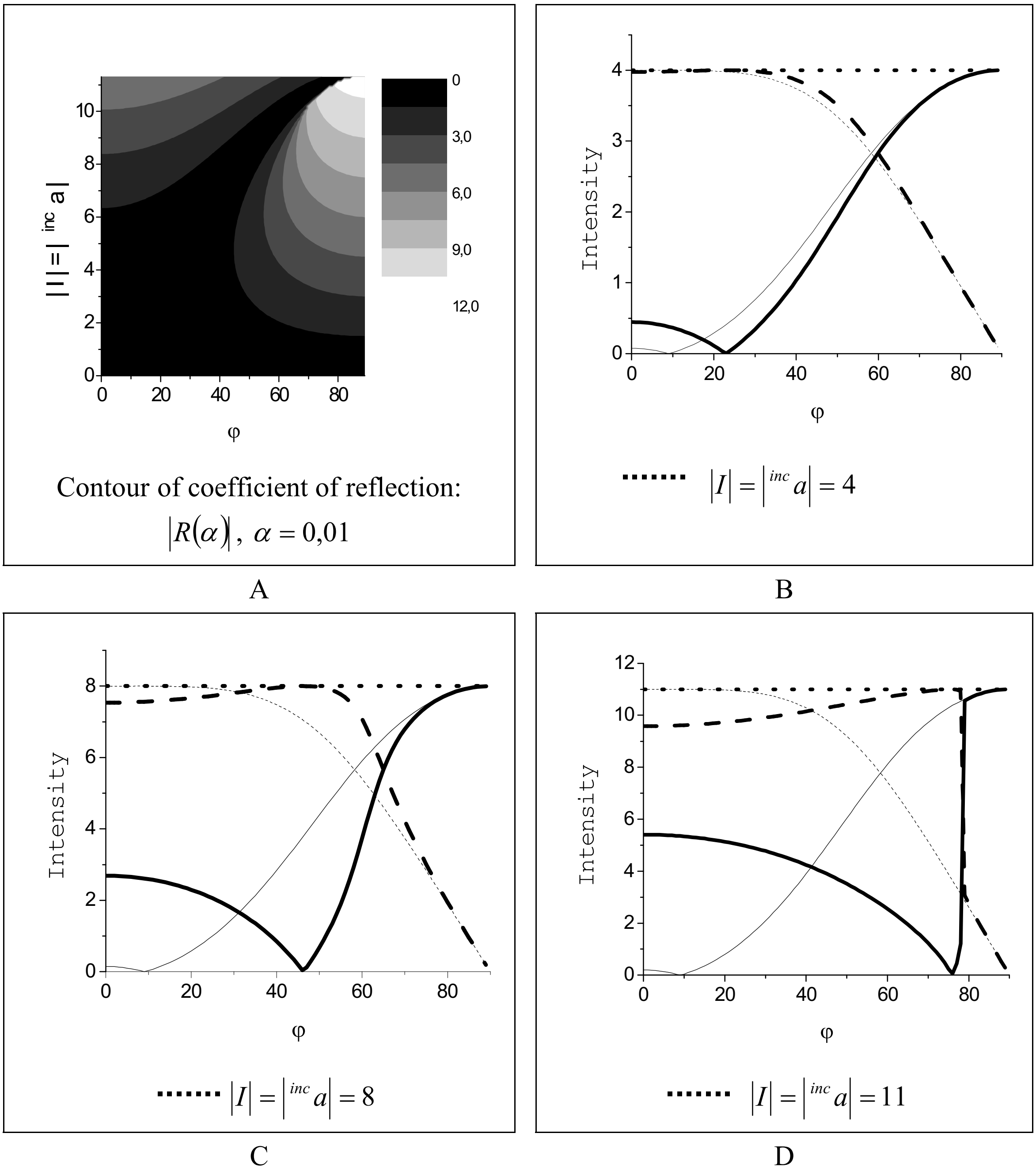

**Figure 4.** Parameters and designations for nonlinear ($\alpha = 0{,}01$) and linear ($\alpha \equiv 0$) layer:

$$\varepsilon^{(L)} = 16;\ \delta = 0{,}5;\ \kappa = 0{,}5;$$

——— $|R(\alpha)|$ and – – – $|T(\alpha)|$ for **nonlinear layer** with $\alpha = 0{,}01$;

——— $|R(\alpha)|$ and ------ $|T(\alpha)|$ for **linear layer** with $\alpha \equiv 0$.

These effects (see sections 3.1 and 3.2) are connected to resonant properties of a nonlinear dielectric layer and caused by increase of a variation of dielectric permeability of a layer (its nonlinear components) at increase of intensity of a field of excitation of researched nonlinear object, see Fig. 3.

The given results of calculations are received with use of the iteration scheme (10). In a considered range of a variation of parameters of a nonlinear problem of diffraction systems of the equations of dimension were used $N = 101$. Also the relative size of an error was set $\xi = 10^{-7}$.

Algorithms (11)-(16) and (17)-(23) decisions of the nonlinear integrated equation (8) (the nonlinear problem of diffraction (4)-(6)) have a number of the advantages peculiar to a method of Newton. They may be used at the analysis of processes of interaction of waves and also when growth of intensity of a field of excitation results in infringement of convergence of the iteration scheme (10).

# 4 Conclusion

The proposed algorithms and results of the numerical analysis are applied: at investigation of processes of wave self-influence [2]; at the analysis of amplitude-phase dispersion of eigen oscillation-wave fields in the nonlinear objects (the norm problem of own field of a nonlinear structure) [13]; development of the approach of the description of evolutionary processes near to critical points of the amplitude-phase dispersion of nonlinear structure (the case of a linear problem in [14, 15] is considered).